# Encapsulation theory: the transformation equations of absolute information hiding.

## Edmund Kirwan[*]

www.EdmundKirwan.com

## Abstract

*This paper describes how the maximum potential number of edges of an encapsulated graph varies as the graph is transformed, that is, as nodes are created and modified. The equations governing these changes of maximum potential number of edges caused by the transformations are derived and briefly analysed.*

## 1. Introduction

The maximum potential number of edges (M.P.E.) of an encapsulated graph was introduced in [1], which derived the equations for the M.P.E. of any given encapsulated graph of absolute information hiding. These equations, however, were static, offering no insight into the evolution of a graph over time.

This paper addresses this evolutionary aspect by deriving the equations which describe not the overall M.P.E. of a graph but the changes in M.P.E. as a graph undergoes an arbitrary series of transformations.

This paper considers encapsulated graphs of absolute information hiding only.

## 2. Standard deviation

Before examining the transformation equations themselves, let us peform some experiments whose results we shall compare with those we might intuitively expect.

Theorem 1.11 in [1] showed that, given two otherwise equivalent encapsulated graphs, the graph whose information hidden nodes are unevenly distributed over encapsulated regions can never have an M.P.E. of less than that of the graph with evenly distributed information hidden nodes.

This may be understood qualitatively by considering the internal M.P.E. of an encapsulated region which, as also shown in [1], was shown to be proportional to the square of the number of nodes in that region. Thus consider an encapsulated graph of evenly distributed nodes where each encapsulated region has 10 nodes; each region will have an internal M.P.E. of 90 ($=10^2 - 10$). A node moved from one encapsulated region to another will (in a sense we shall later define precisely) increase the, "Unevenness," of the distribution: now one region will have 11 nodes and an M.P.E. of 110 ($=11^2 - 11$), whereas the donor region will have an M.P.E. of 72 ($=9^2 - 9$): moving this node has caused an overall net M.P.E. increase of 2.

Loosely speaking, being proportional to the square of the number of nodes in a region, the internal M.P.E. tends to amplify deviations from even distribution, so the more unevenly distributed a graph is, the greater its M.P.E.

Can we establish a more formal basis for investigation this relationship? Can we rigorously measure this, "Unevenness?"

Indeed we can, by using a tool of the statistician: the standard deviation. The standard deviation measures how widely spread the values in a dataset are. We shall use it first to measure how widely spread the number of information hidden nodes per encapsulated region is, that is, to measure the hidden node distribution.

If we take a graph of $r$ encapsulated regions where $x_i$ is the number of hidden nodes per region and where $\bar{x}$ is the average number of hidden nodes per region, then the standard deviation is defined by the equation:

$$\delta = \sqrt{\frac{1}{r} \sum_{i=1}^{r} \left( x_i - |\bar{x}| \right)^2}$$

The standard deviation of the hidden node distribution for an evenly distributed encapsulated graph is 0; this figure then rises as the graph becomes increasingly unevenly distributed.

Instead of examing how the M.P.E. behaves as the standard deviation of the hidden node distribution increases, however, it is useful to instead examine how the isoledensal configuration efficiency (also defined in [1]) behaves, as the configuration efficiency, being defined between 0 and 1, helps to normalise the trend for graphs of different cardinalities. Thus, whereas we expect that the M.P.E. of a graph will rise as the standard deviation of its hidden node distribution increases, we expect the configuration efficiency of that graph to fall as its standard deviation increases.

Finally, we need only state the actual means of increasing the unevenness of a graph distribution. We shall begin, not with a perfectly evenly distributed graph, but with an graph of, say, 100 encapsulated regions, each region having one information hiding violational node, and a random number – between 0 and 30 – of information hidden nodes. Being thus unevenly distributed, the the graph will have a standard deviation of hidden node distribution of some non-zero number. We shall then take one hidden node from a region and move it to an arbitrarily designated target region. We shall then record the change in the hidden node distribution standard deviation and its resulting change in configuration efficiency. We shall then move a second hidden node from a region into the target region and perform the measurement again. This shall be repeated until all the hidden nodes of the graph are in the target region, thus maximising the hidden node distribution standard deviation.

Figure 1 shows the resulting configuration efficiency plotted as a function of the changing hidden node distribution standard deviation.

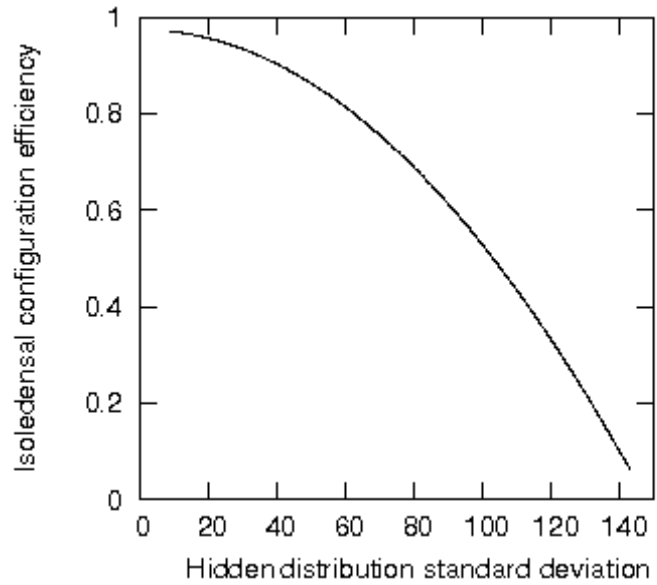

*Figure 1: Isoledensal configuration efficiency as a function of increasing standard deviation of the hidden node distribution.*

This figure shows the expected result: the isoledensal configuration efficiency falls as the hidden node distribution standard deviation increases, i.e., as the encapsulated graph becomes increasingly unevenly distributed in hidden nodes.

To show a slightly broader example of this trend, Figure 2 shows a further ten graphs randomly-generated but with the same constraints as that in figure 1.

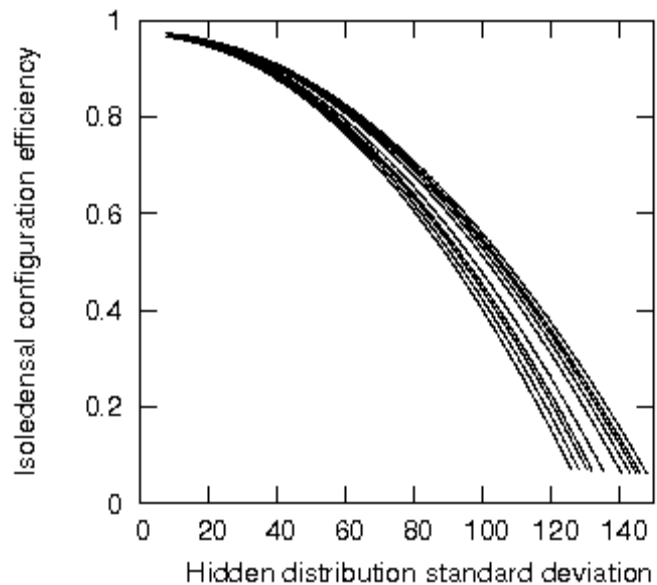

*Figure 2: Isoledensal configuration efficiency as a function of increasing standard deviation, multiple graphs.*

We shall now examine the five transformation equations and attempt to confirm the above result in terms of the appropriate transformation.

# 3. The transformation equations

From the point of view of invesigating changes in M.P.E. there are only two fundamental transformations we can make to any encapsulated graph: we can add a given number of information hiding violational nodes to an encapsulated region of the graph and we can add a given number of information hidden nodes to an encapsulated region of the graph. Let us denote this, "Given number," referenced above, $m$.

It may seem that we could also transform a graph by adding an encapsulated region itself, but in encapsulation theory edges can only be formed between nodes, not regions, and so adding any number of empty regions to a graph cannot change the M.P.E. of that graph. Of course, by definition, no node can exist outside an encapsulated region, these transformations therefore presume the existence of an empty encapsulated region into which new nodes may be introduced, where necessary.

Although only two transformations are fundamental, it is possible to derive a further three transformations from these two fundamental transformation. These three derived transformations both cover common changes to encapsulated graphs and yield important insight into the nature of the changing M.P.E. in their own right. The three derived transformations are: moving $m$ information hiding violational nodes from one encapsulated region to another, moving $m$ information hidden nodes from one encapsulated region to another, and converting $m$ information hidden nodes in an encapsulated region into $m$ information hiding violational nodes.

We note that $m$ may be negative and we establish the convention that the addition of a negative number of nodes to an encapsulated region may be interpreted as the removal of $|m|$ nodes from that region. Where $m$ is negative, it may not exceed the number of nodes that actually reside within an encapsulated region: no region may contain a negative number of nodes at any time.

Before proceeding, we recall the definitions of the terms from [1]:

$G$: An encapsulated graph

$s(G)$: The M.P.E. of encapsulated graph $G$.

$\Delta s(G)$ : The change of M.P.E. of encapsulated graph $G$.

$\left| h(G) \right|$ : The number of information hiding violational nodes in encapsulated graph $G$.

$K_x$, $K_s$ and $K_t$: Encapsulated regions in encapsulated graph $G$. $K_x$ is used when only one encapsulated region is involved. Translation transformations involve two encapsulated regions: $K_s$ is the source region from which nodes are moved, $K_t$ is the target region to which nodes are moved.

$\left| K_x \right|$ : The total number of nodes in encapsulated region $K_x$.

$\left| h(K_x) \right|$ : The number of information hiding violational nodes in encapsulated region $K_x$.

$n$: the total number of nodes in an encapsulated graph.

Table 1 lists the five transformation equations.

| Equation | Description |
|---|---|
| $\Delta s(G) = m(n + |K_x| + |h(G)| - |h(K_x)| + m - 1)$ | The information hiding violation transformation equation (see theorem 3.11), which gives the change of M.P.E. of an encapsulated graph $G$ when $m$ information hiding violational nodes are added to encapsulated region $K_x$. |
| $\Delta s(G) = m(|h(G)| - |h(K_x)| + 2|K_x| + m - 1)$ | The information hidden transformation equation (see theorem 3.17), which gives the change of M.P.E. of an encapsulated graph $G$ when $m$ information hidden nodes are added to encapsulated region $K_x$. |
| $\Delta s_{cumulative}(G) = m(|K_t| - |h(K_t)| - (|K_s| - |h(K_s)|))$ | The information hiding violation translation transformation equation (see theorem 3.18), which gives the change of M.P.E. of an encapsulated graph $G$ when $m$ information hiding violational nodes are moved from source encapsulated region $K_s$ to target encapsulated region $K_t$. |
| $\Delta s_{cumulative}(G) = m(2|K_t| - 2|K_s| + |h(K_s)| - |h(K_t)| + 2m)$ | The information hidden translation transformation equation (see theorem 3.19), which gives the change of M.P.E. of an encapsulated graph $G$ when $m$ information hidden nodes are moved from source encapsulated region $K_s$ to target encapsulated region $K_t$. |
| $\Delta s_{cumulative}(G) = m(n - |K_x|)$ | The conversion transformation equation (see theorem 3.20), which gives the change of M.P.E. when $m$ information hidden nodes are converted into information hiding violational nodes. |

*Table 1: The five transformation equations.*

# 4. Reflections on the equations

## 4.1 The non-conservative transformation equations

Consider the first two transformation equations. These are the fundamental equations and they are non-conservative in that they change the total number of nodes in the graph; the other three equations are conservative in that they do not change the total number of nodes in the graph.

Perhaps the most important aspect of the two non-conservative transformation equations is that it is trivial to show (by subtracting the second from the first) that adding a violational node to a graph causes a larger increase in M.P.E. than adding a hidden node, as we intuitively expect.

## *4.2 The translation transformation equations*

The third and fourth equations are derived from the first two. These equations are translation equations in that they show how M.P.E. changes as nodes are translated or moved from one encapsulated region to another. We shall examine them in reverse order.

### 4.2.1 The fourth equation

Consider the fourth transformation equation, describing the change of M.P.E. as information hidden nodes are moved between encapsulated regions. This is the equation governing the changes that we found in section 2, where all the hidden nodes of a graph were incrementally translated from their original encapsulated regions to a specific target encapsulated region, thereby maximising the standard deviation of the hidden node distribution.

If we look at the terms of the fourth equation, we see that there are three components of the M.P.E. change (ignoring the common scaling *m* factor):

(i) $2|K_t| - 2|K_s|$

(ii) $|h(K_s)| - |h(K_t)|$

(iii) $2m$

The *2m* component is clearly independent of the encapsulated regions affected by the transformation.

Component (i) is the difference in the total number of nodes (multiplied by two) between the source and target encapsulated regions.

Component (ii) is the difference in the number of information hiding violational nodes between the source and target encapsulated regions, though in the opposite sense of component (i) in that component (i) is target minus source but component (ii) is source minus target.

The interaction between these two components is complicated, but in our experiment in section 2, we moved more and more hidden nodes into a single, target encapsulated region, causing the target region to become increasingly large while its violational nodes remained unchanged: thus component (i) grew to dominate component (ii) and repeated translations increasingly added to the M.P.E. of the graph. Increasing the M.P.E. of a fixed number of nodes by definition decreases the graph's configuration efficiency and this is precisely what the graph in figures 1 and 2 show.

The reverse is also true: moving information hidden nodes from a larger to a smaller encapsulated region must necessarily decrease the M.P.E. of a graph and thus increase the configuration efficiency. This explains why a graph of uniformly distributed hidden nodes cannot have an M.P.E. greater than one of unevenly distributed hidden nodes (information hiding violation distribution being equal).

### 4.2.2. The third equation

Whereas the fourth transformation moved information hidden nodes between encapsulated regions, the third transformation moves information hiding violational nodes between encapsulated regions.

We might expect this equation to yield results quite similar to the fourth equation given that they are both translation transformations, but this is not the case. To appreciate this curious difference we shall perform another experiment.

In section 2 we plotted the falling configuration efficiency of a graph as its hidden nodes were increasinly piled into just one encapsulated region. Let us perform a similar experiment but this time we shall incrementally move only the information hiding violational nodes into one encapsulated region. (A minor difference in procedure must be observed: every encapsulated region must contain at least one violational node as otherwise it is uncontactable by nodes in other regions, so instead of moving all violation nodes from the source regions, we shall move all but one violation node. This difference in itself should not materially change the outcome.)

Let us again take an encapsulated graph of 100 encapsulated regions. In section 2 we put one information hiding violational node in each region and put a random number – between 0 and 30 – of information hidden nodes in each region. For our new experiment we shall do the opposite, putting one information hidden node in each region and puting a random number – between 0 and 30 – of information hiding violational nodes in each region.

In section 2, it was the stanard deviation of the hidden node distribution that we measured; this time we shall measure the standard deviation of the violational node distribution: the number of information hiding violational nodes per region. Incrementally moving one violational node into one target region will continuously increase the standard deviation of the violational node distribution of the entire graph, exactly analogous to the first experiment. The question is: how will the configuration efficiency change as the standard deviation of the violational node distribution increases? The result is shown in figure 3.

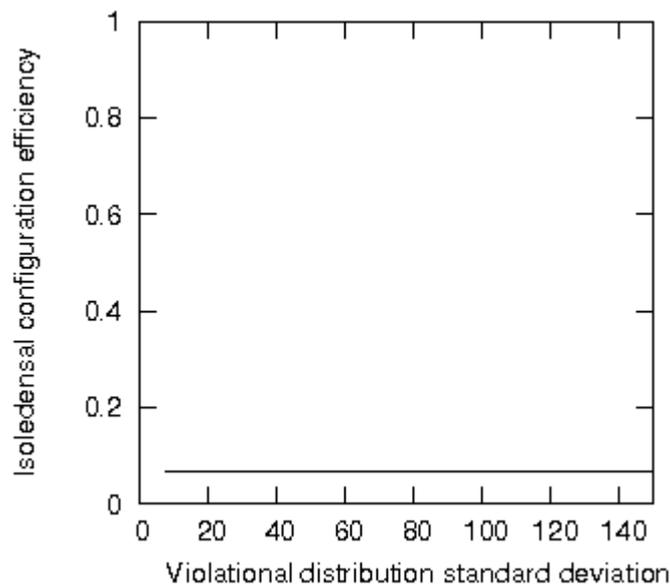

*Figure 3: Isoledensal configuration efficiency as a function of increasing standard deviation of the violational node distribution.*

Figure 3 shows two things but only one is relevant here. Firstly, the low the configuration efficiency of this graph may be striking but this is simply due to the way we created the graph. The graph was created with just one hidden node in each region and up to 30 violational nodes in each region; the graph is therefore composed overwhelming of violational nodes and as such exploits little encapsulation: hence the low configuration efficiency.

More relevant to our experiment and perhaps more surprising is the result that in our sample graph the configuration efficiency is independent of violational distribution standard deviation; in other words, the M.P.E. of the graph is unchanged by moving all the possible violational nodes into one encapsulated region.

Why is this result so different from the first experiment with information hidden nodes?

The answer lies in the third equation.

Let us again break the equation into its component parts; we see it has just two (ignoring the common scaling $m$ factor) :

(i)    $\left|K_t\right|-\left|h\left(K_t\right)\right|$

(ii)    $\left|K_s\right|-\left|h\left(K_s\right)\right|$

Looking at the first component, we may ask ourselves: how do we interpret this quantity? This component is the total number of nodes in the target encapsulated region minus the number of violational nodes in the target region. We are already familiar with this quantity, however: it all the violational nodes are removed from a region, all this is left are the information hidden nodes. Thus the first component is simply the number of hidden nodes in the target region. The second component is just the number of hidden nodes in the source region.

This explains figure 3. Moving violational nodes between regions depends only on the difference between the number of hidden nodes of the source and target region, and in our experiment all the regions have the same number of hidden nodes, thus the difference between the number of hidden nodes in any two regions is zero. So it doesn't matter how many violational nodes are moved between regions: these translations cannot change the M.P.E. and cannot change the configuration efficiency of the graph.

### 4.2.3. Equal unevenness

In both experiments performed so far, one set of nodes were always evenly distributed and minimised: in the first experiment, each region had just one violational node whereas the hidden nodes were unevenly distributed; in the second experiment, each region had one hidden node whereas the violational nodes were unevenly distributed. To model more, "Real world," problems, we must examine graphs whose hidden nodes and violational nodes are both unevenly distributed.

Let us re-visit the first experiment and look at the translation of hidden nodes in a graph again of 100 encapsulated regions with each region having a random number – between 0 and 30 – of hidden nodes and a random number – between 1 and 30 – of violational nodes. Before we do so, however, we shall attempt to predict the results by examining the translation transformation equation for hidden nodes, the fourth equation in table 1.

As we noted before, the dominant component of translation transformation equation for hidden nodes is simply the total number of nodes in the target minus the total number of nodes in the source. As we are chosing at random the target region into which all the hidden nodes will be moved, then this region will initially have between 0 and 60 nodes in total (there will be at most 30 hidden and at most 30 violational nodes).

It is therefore conceivable that the first source region chosen for a translation will have more nodes than our target region, and so the total number of nodes in the target region minus the total number of nodes in the source region will be negative; this negative change in M.P.E. implies that the configuration efficiency of the graph would initially rise. This movement of hidden nodes from a larger to a small region, however, also implies that the standard deviation of the hidden node distribution falls, as the graph is being more, "Smoothed-out," by such a translation. Thus a negative change of M.P.E. corresponds to a negative change of standard deviation.

As more translations are performed, however, we should quickly reach a situation, in a randomly distributed graph, where the number of hidden nodes in the target region becomes greater than the number of nodes in any other single region; this will certainly be the case when the target region contains 61 nodes and will probably be the case much sooner. After this point, all the hidden node translations will increase the M.P.E. and increase the standard deviation of the hidden node distribution. This will yield a picture very similar to that already shown in figure 1, where the configuration efficiency will fall as the hidden node distribution standard deviation increases. The results are shown in figure 4.

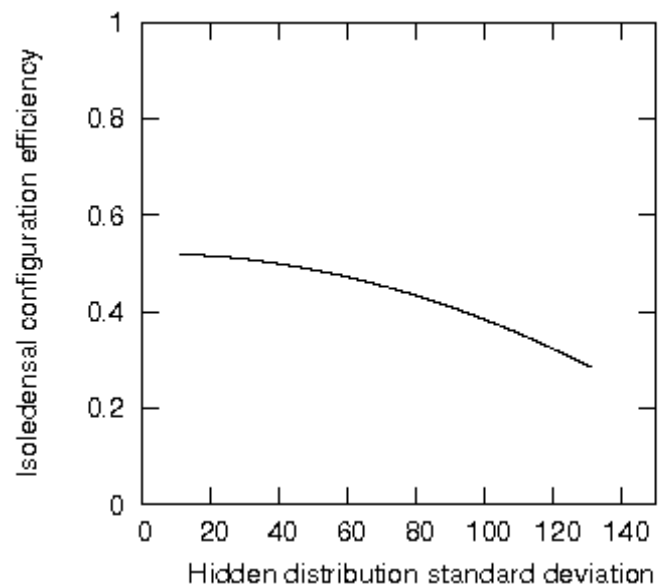

*Figure 4: Isoledensal configuration efficiency as a function of increasing standard deviation of the hidden node distribution for a graph with unevenly distributed hidden and violational nodes.*

There are three aspects of figure 4 that require explanation but only two are relevant to our experiment.

Compared with figure 1, the initial configuration efficiency of the graph in figure 4 is rather low, but this is due to the graph's now containing multiple hidden and multiple violational nodes randomly distributed: any such graph with a non-trivial number of encapsulated regions will usually have a configuration efficiency of around 0.5. (Recall that the graph in figure 1 had only one violational node per region: it was extremely well encapsulated and hence its configuration efficiency was much higher than our latest graph.)

The second and most interesting aspect of figure 4 is that, as suspected, the configuration efficiency of a graph of unevenly distributed hidden and violational nodes falls with increasing hidden node distribution standard deviation, just as was the case with the unevenly distributed hidden nodes and evenly distributed violational nodes of the first experiment.

There only remains to be explained why the terminal configuration efficiency of the graph in figure 4 is not as low as that in figure 1. The explanation comes again from the fourth equation. The largest change of M.P.E. occurs when the difference in total number of nodes between the source and target regions is maximised. In our first experiment, all the regions contained only one violational node but in this latest experiment there were always a random number of violational nodes left behind when the hidden nodes were extracted, thus the differences in total number of nodes between target and source regions were not as great as those in the first experiments causing the M.P.E. to rise by a lesser amount than in the first experiment. This directly translates to the configuration efficiency's not falling as far in our latest experiment.

As before, merely to demonstrate the trend, figure 5 shows ten randomly generated graphs, each constrained as was the graph in figure 4, subjected to repeated hidden translation transformations.

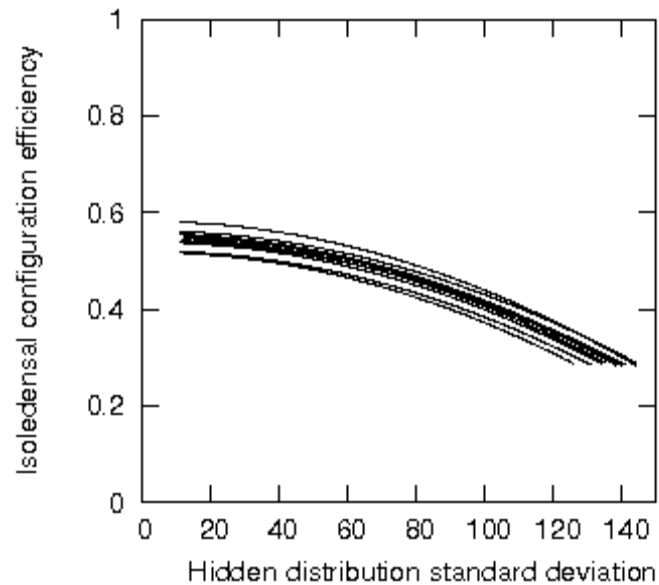

*Figure 5: Isoledensal configuration efficiency as a function of increasing standard deviation of the hidden node distribution for a graph with unevenly distributed hidden and violational nodes, multiple graphs.*

Let us also re-visit the second experiment and look at the translation of violational nodes in a graph again of 100 encapsulated regions with each region having a random number – between 0 and 30 – of hidden nodes and a random number – between 1 and 30 – of violational nodes. Before we do so, however, we shall attempt to predict the results by examining the translation transformation equation for violational nodes, the third equation in table 1.

As we saw, the equation showed us that the change in M.P.E. generated by the translation depends only on the difference between the numbers of hidden nodes in the target and source regions. In our previous violation translation experiment, that difference was zero, and so moving all the violational nodes to one target region had no effect on the graph's M.P.E. or configuration efficiency.

In our randomly generated graph, however, the difference will usually be non-zero and so there will usually be a change of M.P.E. It will not, however, resemble the M.P.E. caused by hidden node translations. In hidden node translations, the change in M.P.E. was proportional to the difference of the number of nodes in the target and source regions, and this change grew increasingly large as the target node grew increasingly large. The very act of translating a hidden node to the target region increased the change in M.P.E. caused by translating all subsequent hidden nodes to the target region.

The third equation exhibits very different behaviour. The change in M.P.E. is proportional to the difference in the number of hidden nodes only and moving violational nodes does not change the number of hidden nodes in source or target region. Thus repeated violational node translations will not generate M.P.E. changes proportional to the increasing size of the target region: the change in M.P.E. is fixed by the choice of source and target regions.

Also, suppose the target region has 15 hidden nodes; then translating violational nodes from a source region with fewer than 15 hidden nodes will cause an increase of M.P.E. and translating violational nodes from a source region with more than 15 nodes will cause a decrease of M.P.E. So unlike the hidden node translations, which after an initial time were guaranteed to only increase M.P.E., violational node translations can lead to small increases of M.P.E. which can then be offset by subsequent small decreases of M.P.E.[1]

In other words, although the increasing violational distribution standard deviation will usually change the configuration efficiency, we do not expect configuration efficiency to change by anything like as much as was caused by the hidden node translations: the configuration efficiency should be quite insenitive to increasing violational distribution standard deviation.

---

1   This justifies the notion of the interface repository in computer programming, a subsystem holding only public
    interfaces that act as facades to various other subsystems of hidden implementations.

The results of repeated violational node translations of a graph unevenly distributed in both hidden and violational nodes is shown in figure 6.

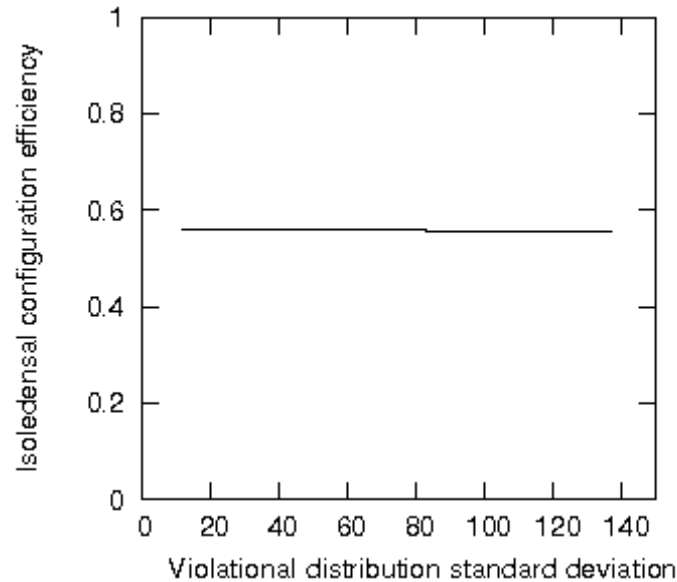

*Figure 6: Isoledensal configuration efficiency as a function of increasing standard deviation of the violation node distribution for a graph with unevenly distributed hidden and violational nodes, multiple graphs.*

As figure 6 indeed shows, configuration efficiency does change with increasing violational distribution standard deviation, but only negligibly. Figure 7 shows the results for multiple graphs.

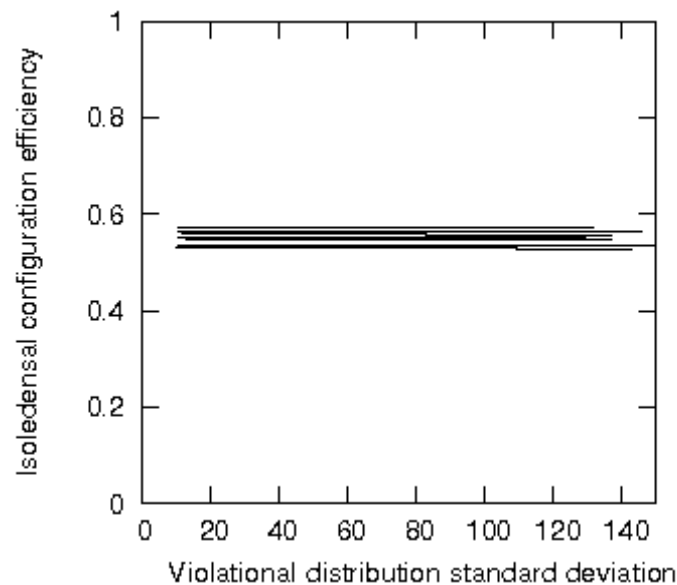

*Figure 7: Isoledensal configuration efficiency as a function of increasing standard deviation of the violation node distribution for a graph with unevenly distributed hidden and violational nodes, multiple graphs.*

## *4.3. The fifth equation*

The fifth equation in table 1 is the conversion transformation equation, which gives the change of M.P.E. when *m* information hidden nodes are converted into information hiding violational nodes. The equation simply states that when a hidden node in a region is converted into a violational node in that same region, then the M.P.E. must rise. This just confirms our expectation that increasing the access to a node within a region must raise the M.P.E. of an encapsulated graph. The reverse is also true: the change of M.P.E. in converting a violational node to a hidden node can be calculated by changing the sign of the *m*, which then shows that such a conversion must reduce the M.P.E. of a graph.

# 5. Conclusions

Systems evolve. To control this evolution means be able to deterministically predict the affects of changes before those changes occur. For systems that can be modelled by encapsulated graphs, this primarily means predicting the M.P.E. of the graph before the changes occur. This paper proposed that the evolution of an encapsulated graph may be modelled as an arbitrarily complex series of transformations that may be applied to that graph.

The two fundamental transformations were then established that describe all changes to a graph and the two M.P.E. equations corresponding to those transformations were proposed. Three further equations were derived from these fundamental equations to describe the more common changes that graphs undergo.

# 6. Related work

-

# 7. Appendix A

## 7.1. Definitions

[D3.1] Given encapsulated region *K* in encapsulated graph *G* and the transformation *T*, the change of internal maximum potential number of edges $\Delta s_{in}(K)$ effected by applying *T* to *K* is given by equation:

$$\Delta s_{in}(K) = s_{in}(T(K)) - s_{in}(K)$$

The change of internal maximum potential number of edges $\Delta s_{in}(G)$ effected by applying *T* to *G* is given by equation:

$$\Delta s_{in}(G) = s_{in}(T(G)) - s_{in}(G)$$

[D3.2] Given encapsulated region *K* in encapsulated graph *G* and the transformation *T*, the change of external maximum potential number of edges $\Delta s_{ex}(K)$ effected by applying *T* to *K* is given by equation:

$$\Delta s_{ex}(K) = s_{ex}(T(K)) - s_{ex}(K)$$

The change of external maximum potential number of edges $\Delta s_{ex}(G)$ effected by applying *T* to *G* is given by equation:

$$\Delta s_{ex}(G) = s_{ex}(T(G)) - s_{ex}(G)$$

[D3.3] Given encapsulated region *K* in encapsulated graph *G* and the transformation *T*, the change of maximum potential number of edges $\Delta s(K)$ effected by applying *T* to *K* is given by equation:

$$\Delta s(K) = s(T(K)) - s(K)$$

The change of maximum potential number of edges $\Delta s(G)$ effected by applying *T* to *G* is given by equation:

$$\Delta s(G) = s(T(G)) - (G)$$

[D3.4] Given the encapsulated graph *G* with an *i*$^{th}$ encapsulated region $K_i$ of $|K_i|$ nodes and external information hiding violation of *h(Ki)*, let $T_h$ be the violational transformation which maps $K_i$ onto $K_i^*$

where $K_i^*$ differs from $K_i$ by $m$ information hiding violational nodes where $m \geq -|K_i|$ or:

$$T_h(K_i, m) = \{K_i \in G : |h(K_i)| \rightarrow |h(K_i)| + m\}$$

Where $T_h$ is applied to just the $x^{th}$ encapsulated region $K_x$ of $G$, the transformation becomes:

$$T_h(x, G, m) = \{G \rightarrow T_h(G) : |h(K_i)| \rightarrow |h(K_i)| + m \, \forall \, i = x \, ; |h(K_i)| \rightarrow |h(K_i)| \, \forall \, i \neq x\}$$

Note that as $m$ may be positive or negative, $K_i^*$ may have more or fewer information hiding violational nodes than $K_i$.

[D3.5] Given the encapsulated graph $G$ with an $i^{th}$ encapsulated region $K_i$ of $|K_i|$ nodes and external information hiding violation of $h(K_i)$, let $T_z$ be the hidden transformation which maps $K_i$ onto $K_i^*$ where $K_i^*$ differs from $K_i$ by $m$ information hidden nodes where $m \geq -|K_i|$ and where the information hiding violational nodes remain unchanged, that is:

$$T_z(K_i, m) = \{K_i \in G : |K_i| \rightarrow |K_i| + m \, ; |h(K_i)| \rightarrow |h(K_i)|\}$$

Where $T_z$ is applied to just the $x^{th}$ encapsulated region $K_x$ of $G$, the transformation becomes:

$$T_z(x, G, m) = \{G \rightarrow T_z(G) : |K_i| \rightarrow |K_i| + m \, \forall \, i = x \, ; |K_i| \rightarrow |K_i| \, \forall \, i \neq x \, ; |h(K_i)| \rightarrow |h(K_i)| \, \forall \, i\}$$

Note that as $m$ may be positive or negative, $K_i^*$ may have more or fewer information hidden nodes than $K_i$.

## 7.2. Theorems

The theorems are organised as follows.

Theorems 3.1 – 3.5 establish some general results concerning the sum of changes of maximum potential number of edges and the application of transformations to graphs.

Theorems 3.6 – 3.11 establish the fundamental transformation equation for the application of the violational transformation to a graph.

Theorems 3.12 – 3.17 establish the fundamental transformation equation for the application of the hidden transformation to a graph.

Theorems 3.18 – 3.20 establish the three derived transformation equations for the translation and conversion transformations.

All theorems relate to encapsulated graphs of absolute information hiding only.

## Theorem 3.1.

Given encapsulated region $G$ and the transformation $T$, the change of maximum potential number of edges $s(G)$ effected by applying $T$ to $G$ is given by:

$$\Delta s(G) = \Delta s_{in}(G) + \Delta s_{ex}(G)$$

*Proof:*

By definition:

$$s(G) = s_{in}(G) + s_{ex}(G) \quad (i)$$

Let $K^* = T(K)$. Therefore:

$$s(G^*) = s_{in}(G^*) + s_{ex}(G^*) \quad (ii)$$

Also be definintion:

$$\Delta s(G) = s(T(G)) - s(G) \quad (iii)$$

Substituting (i) and (ii) into (iii) gives:

$$\Delta s(G) = s(G^*) - s(G)$$
$$= s_{\text{in}}(G^*) + s_{\text{ex}}(G^*) - s_{\text{in}}(G) - s_{\text{ex}}(G)$$
$$= s_{\text{in}}(G^*) - s_{\text{in}}(G) + s_{\text{ex}}(G^*) - s_{\text{ex}}(G)$$
$$= \Delta s_{\text{in}}(G) + \Delta s_{\text{ex}}(G)$$

*QED*

## Theorem 3.2.

Given the encapsulated region *K* and the violational transformation $T_h$ defined in [D3.4], then the number of nodes in $T_h(K)$ differs from the number of nodes in *K* by *m,* or:

$$\left| T_h(K, m) \right| = |K| + m$$

*Proof:*

Let *K* contain *a* information hidden nodes and $\left| h(K) \right|$ information hiding violational nodes. Thus, by definition:

$$|K| = a + \left| h(K) \right| \qquad \text{(i)}$$

Let $K^* = T_h(K, m)$. By definintion $T_h$ leaves the number of information hidden nodes unchanged, therefore:

$$\left| K^* \right| = a + \left| h(K^*) \right| \qquad \text{(ii)}$$

Also by definition of $T_h$:

$$\left| h(K^*) \right| = \left| h(K) \right| + m \qquad \text{(iii)}$$

Substituting (iii) into (ii) gives:

$$\left| K^* \right| = a + \left| h(K) \right| + m \qquad \text{(iv)}$$

Substituting (i) into (iv) gives:

$$\left| T_h(K, m) \right| = \left| K^* \right| = |K| + m$$

*QED*

## Theorem 3.3.

Given the encapsulated graph *G* with an $i^{th}$ encapsulated region $K_i$ and given that a particular $x^{th}$ encapsulated region $K_x$ only is subject to the violational transformation $T_h$ defined in [D3.4], the number of information hiding violation nodes in $T_h(G)$ is given by:

$$\left| h(T_h(x, G, m)) \right| = \left| h(G) \right| + m$$

*Proof:*

By definintion:

$$\left| h(G) \right| = \sum_{i=1}^{r} \left| h(K_i) \right|$$

$$= \sum_{i=1 \neq x}^{r} \left| h\left(K_i\right)\right| + \left| h\left(K_x\right)\right| \qquad \text{(i)}$$

Let $G^* = T_h(x, G, m)$ and let $K_i^* = T_h(K_i, m)$. By definition:

$$\left| h\left(G^*\right)\right| = \sum_{i=1}^{r} \left| h\left(K_i^*\right)\right|$$

$$= \sum_{i=1 \neq x}^{r} \left| h\left(K_i^*\right)\right| + \left| h\left(K_x^*\right)\right| \qquad \text{(ii)}$$

By the definintion of $T_h$:

$$\forall\, i \neq x : \left| h\left(K_i\right)\right| \rightarrow \left| h\left(K_i\right)\right| \qquad \text{(iii)}$$

Substituting (iii) into (ii) gives:

$$\left| h\left(G^*\right)\right| = \sum_{i=1 \neq x}^{r} \left| h\left(K_i\right)\right| + \left| h\left(K_x^*\right)\right| \qquad \text{(iv)}$$

Also by the definintion of $T_h$:

$$\forall\, i = x : \left| h\left(K_i\right)\right| \rightarrow \left| h\left(K_i\right)\right| + m \qquad \text{(v)}$$

Substituting (v) into (iv) gives:

$$\left| h\left(G^*\right)\right| = \sum_{i=1 \neq x}^{r} \left| h\left(K_i\right)\right| + \left| h\left(K_x\right)\right| + m \qquad \text{(vi)}$$

Substituting (i) into (vi) gives:

$$\left| h\left(T_h(x, G, m)\right)\right| = \left| h\left(G^*\right)\right| = \left| h(G)\right| + m$$

<div align="right"><em>QED</em></div>

## Theorem 3.4.

Given the encapsulated region $K$ and the hidden node transformation $T_z$ defined in [D3.5], then the number of nodes in $T_z(K)$ differs from the number of nodes in $K$ by $m$, or:

$$\left| T_z(K, m)\right| = \left| K\right| + m$$

*Proof:*

Let $K$ contain $a$ information hidden nodes and $\left| h(K)\right|$ information hiding violational nodes. Thus, by definition:

$$\left| K\right| = a + \left| h(K)\right| \qquad \text{(i)}$$

Let $K^* = T_z(K, m)$. By definintion $T_z$ leaves the number of information hiding violational nodes unchanged, therefore:

$$\left| K^*\right| = a^* + \left| h(K)\right| \qquad \text{(ii)}$$

Also by definition of $T_z$:

$$a^* = a + m \qquad \text{(iii)}$$

Substituting (iii) into (ii) gives:

$$\left| K^*\right| = a + m + \left| h(K)\right| \qquad \text{(iv)}$$

Substituting (i) into (iv) gives:

$$\left|T_z(K,m)\right|=\left|K^*\right|=\left|K\right|+m$$

*QED*

## Theorem 3.5.

Given the encapsulated graph $G$ with an $i^{th}$ encapsulated region $K_i$ and given that a particular $x^{th}$ encapsulated region $K_x$ only is subject to the hidden node transformation $T_z$ defined in [D3.5], the number of information hiding violation nodes in $T_z(G)$ is given by:

$$\left|h(T_z(x,G,m))\right|=\left|h(G)\right|$$

*Proof:*

Let $G^*=T_z(x,G,m)$ and let $K_i^*=T_z(K_i,m)$. By definition:

$$\left|h(G^*)\right|=\sum_{i=1}^{r}\left|h(K_i^*)\right| \quad \text{(i)}$$

By the definintion of $T_z$:

$$\forall i:\left|h(K_i^*)\right|=\left|h(K_i)\right| \quad \text{(ii)}$$

Substituting (ii) into (i) gives:

$$\left|h(T_z(x,G,m))\right|=\sum_{i=1}^{r}\left|h(K_i)\right|=\left|h(G)\right|$$

*QED*

## Theorem 3.6.

Given the encapsulated graph $G$ with an $i^{th}$ encapsulated region $K_i$ of $\left|K_i\right|$ nodes and internal maximum potential number of edges $s_{in}(K_i)$, the change of $s_{in}(K_i)$ when the number of information hiding violational nodes in $K_i$ changes by $m$ where $m\geq-\left|K_i\right|$ is given by:

$$\Delta s_{in}(K_i)=2m\left|K_i\right|+m^2-m$$

*Proof:*

Let $T_h$ be the violational transformation defined in [D3.4] and let $K_i^*=T_h(K_i,m)$. By definition [D3.1], the change of internal maximum potential number of edges effected by applying $T_h$ to $K_i$ is:

$$\Delta s_{in}(K_i)=s_{in}(K_i^*)-s_{in}(K_i) \quad \text{(i)}$$

By theorem 1.2:

$$s_{in}(K_i^*)=\left|K_i^*\right|(\left|K_i^*\right|-1) \quad \text{(ii)}$$

By theorem 3.2:

$$\left|K_i^*\right|=\left|K_i\right|+m \quad \text{(iii)}$$

Substituting (iii) into (ii) gives:

$$s_{in}(K_i^*)=(\left|K_i\right|+m)(\left|K_i\right|+m-1)$$
$$=\left|K_i\right|^2+m\left|K_i\right|-\left|K_i\right|+m\left|K_i\right|+m^2-m$$

$$= |K_i|(|K_i|-1)+2m|K_i|+m^2-m \quad \text{(iv)}$$

But:

$$s_{\text{in}}(K_i)=|K_i|(|K_i|-1) \quad \text{(v)}$$

Substituting (v) into (iv) gives:

$$s_{\text{in}}(K_i^*)=s_{\text{in}}(K_i)+2m|K_i|+m^2-m$$

$$s_{\text{in}}(K_i^*)-s_{\text{in}}(K_i)=2m|K_i|+m^2-m$$

And therefore by (i):

$$\Delta s_{\text{in}}(K_i)=2m|K_i|+m^2-m$$

*QED*

## Theorem 3.7.

Given the encapsulated graph $G$ with an $i^{th}$ encapsulated region $K_i$ of internal maximum potential number of edges $s_{in}(K_i)$, the change of internal maximum potential number of edges of the entire graph when the number of information hiding violational nodes in a particular $x^{th}$ encapsulated region $K_x$ changes by $m$ where $m \geq -|K_x|$ is equal to the change of internal maximum potential number of edges of $K_x$, or:

$$\Delta s_{\text{in}}(G)=\Delta s_{\text{in}}(K_x)$$

*Proof:*

By definintion, the internal maximum potential number of edges of $G$ is the sum of the internal maximum potential number of edges of all its encapsulated regions:

$$s_{in}(G) = \sum_{i=1}^{r} s_{\text{in}}(K_i)$$

$$= \sum_{i=1 \neq x}^{r} s_{\text{in}}(K_i)+s_{\text{in}}(K_x) \quad \text{(i)}$$

Let $T_h$ be the violational transformation defined in [D3.4] and let $K_i^*=T_h(K_i,m)$. Futhermore, let $T_h$ apply to the $x^{th}$ encapsulated region only such that $K_x^*=T_h(K_x,m)$ and $G^*=T_h(x,G,m)$. By definintion:

$$s_{in}(G^*) = \sum_{i=1}^{r} s_{\text{in}}(K_i^*)$$

$$= \sum_{i=1 \neq x}^{r} s_{\text{in}}(K_i^*)+s_{\text{in}}(K_x^*) \quad \text{(ii)}$$

By definition [D3.3], the change of internal maximum potential number of edges of $G$ effected by applying transformation $T_h$ to G is given by:

$$\Delta s_{\text{in}}(G)=s_{\text{in}}(G^*)-s_{\text{in}}(G) \quad \text{(iii)}$$

Substituting (i) and (ii) into (iii) gives:

$$\Delta s_{\text{in}}(G)=\sum_{i=1 \neq x}^{r} s_{\text{in}}(K_i^*)+s_{\text{in}}(K_x^*)-\sum_{i=1 \neq x}^{r} s_{\text{in}}(K_i)-s_{\text{in}}(K_x) \quad \text{(iv)}$$

But as $T_h$ is only applied to $K_x$ then all encapsulated regions except $K_x$ are unchanged, or:

$$K_i^* = K_i \, \forall \, i \neq x$$

And therefore:

$$s_{\text{in}}(K_i^*) = s_{\text{in}}(K_i) \, \forall \, i \neq x \quad \text{(v)}$$

Substituting (v) into (iv) gives:

$$\Delta s_{\text{in}}(G) = \sum_{i=1 \neq x}^{r} s_{\text{in}}(K_i) + s_{\text{in}}(K_x^*) - \sum_{i=1 \neq x}^{r} s_{\text{in}}(K_i) - s_{\text{in}}(K_x)$$

$$= s_{\text{in}}(K_x^*) - s_{\text{in}}(K_x) \quad \text{(vi)}$$

By definition [D3.1], the change of internal maximum potential number of edges effected by applying $T_h$ to $K_x$ is then:

$$\Delta s_{\text{in}}(K_x) = s_{\text{in}}(K_x^*) - s_{\text{in}}(K_x) \quad \text{(vii)}$$

Substituting (vii) into (vi) gives:

$$\Delta s_{\text{in}}(G) = \Delta s_{\text{in}}(K_x)$$

*QED*

## Theorem 3.8.

Given the encapsulated graph $G$ with an $i^{th}$ encapsulated region $K_i$ of $\left| K_i \right|$ nodes and external maximum potential number of edges $s_{ex}(K_i)$, the change of $s_{ex}(K_i)$ when the number of information hiding violational nodes in $K_i$ changes by $m$ where $m \geq -\left| K_i \right|$ is given by:

$$\Delta s_{\text{ex}}(K_i) = m \left| h(G) \right| - m \left| h(K_i) \right|$$

*Proof:*

Let $T_h$ be the violational transformation defined in [D3.4]; let $K_i^* = T_h(K_n, m)$ and let $G^* = T_h(G, m)$.

By definition [D3.2], the change of external maximum potential number of edges effected by applying $T_h$ to $K_i$ is then:

$$\Delta s_{\text{ex}}(K_i) = s_{\text{ex}}(K_i^*) - s_{\text{ex}}(K_i) \quad \text{(i)}$$

By theorem 1.4:

$$s_{\text{ex}}(K_i^*) = \left| K_i^* \right| \left( \left| h(G^*) \right| - \left| h(K_i^*) \right| \right) \quad \text{(ii)}$$

By theorem 3.3:

$$\left| h(G^*) \right| = \left| h(G) \right| + m \quad \text{(iii)}$$

Substituting (iii) into (ii) gives:

$$s_{\text{ex}}(K_i^*) = \left| K_i^* \right| \left( \left| h(G) \right| + m - \left| h(K_i^*) \right| \right) \quad \text{(iv)}$$

By theorem 3.2:

$$\left| K_i^* \right| = \left| K_i \right| + m \quad \text{(v)}$$

Substituting (v) into (iv) gives:

$$s_{\text{ex}}(K_i^*) = \left| K_i + m \right| \left( \left| h(G) \right| + m - \left( \left| h(K_i) \right| + m \right) \right)$$

$$= \left| K_i + m \right| \left( \left| h(G) \right| + m - \left| h(K_i) \right| - m \right)$$

$$= |K_i + m|(|h(G)| - |h(K_i)|)$$

$$= |K_i||h(G)| - |K_i||h(K_i)| + m|h(G)| - m|h(K_i)|$$

$$= |K_i|(|h(G)| - |h(K_i)|) + m|h(G)| - m|h(K_i)| \quad \text{(vi)}$$

But:

$$s_{\text{ex}}(K_i) = |K_i|(|h(G)| - |h(K_i)|) \quad \text{(vii)}$$

Substituting (vii) into (vi) gives:

$$s_{\text{ex}}(K_i^*) = s_{\text{ex}}(K_i) + m|h(G)| - m|h(K_i)|$$

$$s_{\text{ex}}(K_i^*) - s_{\text{ex}}(K_i) = m|h(G)| - m|h(K_i)|$$

And therefore by (i):

$$\Delta s_{\text{ex}}(K_i) = m|h(G)| - m|h(K_i)|$$

*QED*

## Theorem 3.9.

Given the encapsulated graph $G$ of $n$ nodes with an $i^{th}$ encapsulated region $K_i$ of $|K_i|$ nodes and external maximum potential number of edges $s_{ex}(K_i)$, the change of external maximum potential number of edges of the entire graph when the number of information hiding violational nodes in a particular $x^{th}$ encapsulated region $K_x$ changes by $m$ where $m \geq -|K_x|$ is given by:

$$\Delta s_{\text{ex}}(G) = mn - m|K_x| + \Delta s_{\text{ex}}(K_x)$$

*Proof:*

By definintion, the external maximum potential number of edges of $G$ is the sum of the external maximum potential number of edges of all its encapsulated regions:

$$s_{ex}(G) = \sum_{i=1}^{r} s_{\text{ex}}(K_i)$$

$$= \sum_{i=1 \neq x}^{r} s_{\text{ex}}(K_i) + s_{\text{ex}}(K_x) \quad \text{(i)}$$

Let $T_h$ be the violational transformation defined in [D3.4] and let $K_i^* = T_h(K_i, m)$. Futhermore, let $T_h$ apply to the $x^{th}$ encapsulated region only such that $K_x^* = T_h(K_x, m)$ and $G^* = T_h(x, G, m)$. By definintion:

$$s_{ex}(G^*) = \sum_{i=1}^{r} s_{\text{ex}}(K_i^*)$$

$$= \sum_{i=1 \neq x}^{r} s_{\text{ex}}(K_i^*) + s_{\text{ex}}(K_x^*) \quad \text{(ii)}$$

By definintion [D3.2]:

$$\Delta s_{\text{ex}}(G) = s_{\text{ex}}(G^*) - s_{\text{ex}}(G) \quad \text{(iii)}$$

Substituting (i) and (ii) into (iii) gives:

$$\Delta s_{\text{ex}}(G) = \sum_{i=1 \neq x}^{r} s_{\text{ex}}(K_i^*) + s_{\text{ex}}(K_x^*) - \sum_{i=1 \neq x}^{r} s_{\text{ex}}(K_i) - s_{\text{ex}}(K_x) \quad \text{(iv)}$$

By definintion [D3.2]:

$$\Delta s_{ex}(K_x) = s_{ex}(K_x^*) - s_{ex}(K_x) \qquad (v)$$

Substituting (v) into (iv) gives:

$$s_{ex}(G) = \sum_{i=1 \neq x}^{r} s_{ex}(K_i^*) - \sum_{i=1 \neq x}^{r} s_{ex}(K_i) + \Delta s_{ex}(K_x) \qquad (vi)$$

By theorem 1.4:

$$s_{ex}(K_i^*) = |K_i^*| \left( \left| h(G^*) \right| - \left| h(K_i^*) \right| \right) \qquad (vii)$$

If we consider $K_i$ where $i \neq x$ then as $T_h$ is applied only to $K_x$ then:

$$\left| K_i^* \right| = \left| K_i \right| \forall i \neq x \qquad (viii)$$

Substituting (viii) into (vii) gives:

$$s_{ex}(K_i^*) = |K_i| \left( \left| h(G^*) \right| - \left| h(K_i) \right| \right) \qquad (ix)$$

By theorem 3.3:

$$\left| h(G^*) \right| = \left| h(G) \right| + m \qquad (x)$$

Substituting (x) into (ix) gives:

$$s_{ex}(K_i^*) = |K_i| \left( \left| h(G) \right| + m - \left| h(K_i) \right| \right)$$
$$= |K_i| \left| h(G) \right| - |K_i| \left| h(K_i) \right| + m |K_i|$$
$$= |K_i| \left( \left| h(G) \right| - \left| h(K_i) \right| \right) + m |K_i| \qquad (xi)$$

But:

$$s_{ex}(K_i) = |K_i| \left( \left| h(G) \right| - \left| h(K_i) \right| \right) \qquad (xii)$$

Substituting (xi) into (xii) gives:

$$s_{ex}(K_i^*) = s_{ex}(K_i) + m |K_i|$$

As this holds $\forall i \neq x$ we can take the sum over all encapsulated regions except $x$:

$$\sum_{i=1 \neq x}^{r} s_{ex}(K_i^*) = \sum_{i=1 \neq x}^{r} s_{ex}(K_i) + \sum_{i=1 \neq x}^{r} m |K_i|$$
$$= \sum_{i=1 \neq x}^{r} s_{ex}(K_i) + m \sum_{i=1 \neq x}^{r} |K_i| \qquad (xiii)$$

By definintion,

$$n = \sum_{i=1}^{r} |K_i| = \sum_{i=1 \neq x}^{r} |K_i| + |K_x|$$

So:

$$n - |K_x| = \sum_{i=1 \neq x}^{r} |K_i| \qquad (xiv)$$

Substituting (xiv) into (viii) gives:

$$\sum_{i=1\neq x}^{r} s_{\text{ex}}(K_i^*) = \sum_{i=1\neq x}^{r} s_{\text{ex}}(K_i) + m(n - |K_x|)$$

$$\sum_{i=1\neq x}^{r} s_{\text{ex}}(K_i^*) = \sum_{i=1\neq x}^{r} s_{\text{ex}}(K_i) + mn - m|K_x| \qquad \text{(xv)}$$

Substituting (xv) into (vi) gives:

$$\Delta s_{\text{ex}}(G) = \sum_{i=1\neq x}^{r} s_{\text{ex}}(K_i) + mn - m|K_x| - \sum_{i=1\neq x}^{r} s_{\text{ex}}(K_i) + \Delta s_{\text{ex}}(K_x)$$

$$= mn - m|K_x| + \Delta s_{\text{ex}}(K_x)$$

*QED*

## Theorem 3.10.

Given the encapsulated graph $G$ of $n$ nodes with an $i^{th}$ encapsulated region $K_i$ of $|K_i|$ nodes and external maximum potential number of edges $s_{ex}(K_i)$, the change of external maximum potential number of edges of the entire graph $s_{ex}(G)$ when the number of information hiding violational nodes in a particular $x^{th}$ encapsulated region $K_x$ changes by $m$ where $m \geq -|K_x|$ is given by:

$$\Delta s_{\text{ex}}(G) = mn - m|K_x| + m|h(G)| - m|h(K_x)|$$

*Proof:*

Let $T_h$ be the violational transformation defined in [D3.4] and let it apply to the $x^{th}$ encapsulated region only. By theorem 3.8, the change of external maximum potential number of edges of $K_x$ by the application of $T_h$ to $K_x$ is given by:

$$\Delta s_{\text{ex}}(K_x) = m|h(G)| - m|h(K_x)| \qquad \text{(i)}$$

By theorem 3.9, the change of external maximum potential number of edges of the entire graph $G$ by the application of $T_h$ to $K_x$ is given by:

$$\Delta s_{\text{ex}}(G) = mn - m|K_x| + \Delta s_{\text{ex}}(K_x) \qquad \text{(ii)}$$

Substituting (ii) into (i) gives:

$$\Delta s_{\text{ex}}(G) = mn - m|K_x| + m|h(G)| - m|h(K_x)|$$

*QED*

## Theorem 3.11.

Given the encapsulated graph $G$ of $n$ nodes with an $i^{th}$ encapsulated region $K_i$ of $|K_i|$ nodes, the change of maximum potential number of edges of the entire graph $s(G)$ when the number of information hiding violational nodes in a particular $x^{th}$ encapsulated region $K_x$ changes by $m$ where $m \geq -|K_x|$ is given by:

$$\Delta s(G) = mn + m|K_x| + m|h(G)| - m|h(K_x)| + m^2 - m$$

*Proof:*

Let $T_h$ be the violational transformation defined in [D3.4] and let $K_i^* = T_h(K_i, m)$. Futhermore, let $T_h$ apply to the $x^{th}$ encapsulated region only such that $K_x^* = T_h(K_x, m)$ and $G^* = T_h(x, G, m)$. From theorem 3.7, when the number of information hiding violational nodes in $K_x$ changes by $m$, the change of internal maximum potential number of edges of the entire graph is given by:

$$\Delta s_{\text{in}}(G) = \Delta s_{\text{in}}(K_x) \quad \text{(i)}$$

By theorem 3.6, when the number of information hiding violational nodes in $K_x$ changes by $m$, the change of internal maximum potential number of edges of $K_x$ is given by

$$\Delta s_{\text{in}}(K_x) = 2m|K_x| + m^2 - m \quad \text{(ii)}$$

Substituting (ii) into (i) gives:

$$\Delta s_{\text{in}}(G) = 2m|K_x| + m^2 - m \quad \text{(iii)}$$

From theorem 3.10, when the number of information hiding violational nodes in $K_x$ changes by $m$, the change of exteral maximum potential number of edges is given by:

$$\Delta s_{\text{ex}}(G) = mn - m|K_x| + m|h(G)| - m|h(K_x)| \quad \text{(iv)}$$

By theorem 3.1, the change of maximum potential number of edges of $G$ is given by:

$$\Delta s(G) = \Delta s_{\text{in}}(G) + \Delta s_{\text{ex}}(G) \quad \text{(v)}$$

Substituting (iii) and (iv) into (v) gives:

$$\Delta s(G) = mn - m|K_x| + m|h(G)| - m|h(K_x)| + 2m|K_x| + m^2 - m$$

$$= mn + m|K_x| + m|h(G)| - m|h(K_x)| + m^2 - m$$

*QED*

## Theorem 3.12.

Given the encapsulated graph $G$ with an $i^{th}$ encapsulated region $K_i$ of $|K_i|$ nodes and internal maximum potential number of edges $s_{in}(K_i)$, the change of $s_{in}(K_i)$ when the number of information hidden nodes in $K_i$ changes by $m$ where $m \geq -|K_i|$ is given by:

$$\Delta s_{\text{in}}(K_i) = 2m|K_i| + m^2 - m$$

*Proof:*

Let $T_z$ be the hidden transformation defined in [D3.5] and let $K_i^* = T_z(K_i, m)$. By definition [D3.1], the change of internal maximum potential number of edges effected by applying $T_z$ to $K_i$ is:

$$\Delta s_{\text{in}}(K_i) = s_{\text{in}}(K_i^*) - s_{\text{in}}(K_i) \quad \text{(i)}$$

By theorem 1.2:

$$s_{\text{in}}(K_i^*) = |K_i^*|(|K_i^*| - 1) \quad \text{(ii)}$$

By theorem 3.4:

$$|K_i^*| = |K_i| + m \quad \text{(iii)}$$

Substituting (iii) into (ii) gives:

$$s_{\text{in}}(K_i^*) = (|K_i| + m)(|K_i| + m - 1)$$

$$= |K_i^2| + m|K_i| - |K_i| + m|K_i| + m^2 - m$$

$$= |K_i|(|K_i| - 1) + 2m|K_i| + m^2 - m \quad \text{(iv)}$$

But:

$$s_{\text{in}}(K_i) = |K_i|(|K_i| - 1) \qquad (v)$$

Substituting (v) into (iv) gives:

$$s_{\text{in}}(K_i^*) = s_{\text{in}}(K_i) + 2\text{m}|K_i| + m^2 - m$$

$$s_{\text{in}}(K_i^*) - s_{\text{in}}(K_i) = 2\text{m}|K_i| + m^2 - m$$

And therefore by (i):

$$\Delta s_{\text{in}}(K_i) = 2\text{m}|K_i| + m^2 - m$$

<div align="right"><em>QED</em></div>

## Theorem 3.13.

Given the encapsulated graph $G$ with an $i^{th}$ encapsulated region $K_i$ of internal maximum potential number of edges $s_{in}(K_i)$, the change of internal maximum potential number of edges of the entire graph when the number of information hidden nodes in a particular $x^{th}$ encapsulated region $K_x$ changes by $m$ where $m \geq -|K_x|$ is equal to the change of internal maximum potential number of edges of $K_x$, or:

$$\Delta s_{\text{in}}(G) = \Delta s_{\text{in}}(K_x)$$

*Proof:*

By definintion, the internal maximum potential number of edges of $G$ is the sum of the internal maximum potential number of edges of all its encapsulated regions:

$$s_{in}(G) = \sum_{i=1}^{r} s_{\text{in}}(K_i)$$

$$= \sum_{i=1 \neq x}^{r} s_{\text{in}}(K_i) + s_{\text{in}}(K_x) \qquad (i)$$

Let $T_z$ be the hidden transformation defined in [D3.5] and let $K_i^* = T_h(K_i, m)$. Futhermore, let $T_z$ apply to the $x^{th}$ encapsulated region only such that $K_x^* = T_h(K_x, m)$ and $G^* = T_z(x, G, m)$. By definintion:

$$s_{in}(G^*) = \sum_{i=1}^{r} s_{\text{in}}(K_i^*)$$

$$= \sum_{i=1 \neq x}^{r} s_{\text{in}}(K_i^*) + s_{\text{in}}(K_x^*) \qquad (ii)$$

By definition [D3.1], the change of internal maximum potential number of edges of $G$ effected by applying transformation $T_z$ to $G$ is:

$$\Delta s_{\text{in}}(G) = s_{\text{in}}(G^*) - s_{\text{in}}(G) \qquad (iii)$$

Substituting (i) and (ii) into (iii) gives:

$$\Delta s_{\text{in}}(G) = \sum_{i=1 \neq x}^{r} s_{\text{in}}(K_i^*) + s_{\text{in}}(K_x^*) - \sum_{i=1 \neq x}^{r} s_{\text{in}}(K_i) - s_{\text{in}}(K_x) \qquad (iv)$$

But as $T_z$ is only applied to $K_x$ then all encapsulated regions except $K_x$ are unchanged, or:

$$K_i^* = K_i \: \forall \: i \neq x$$

And therefore:

$$s_{\text{in}}(K_i^*) = s_{\text{in}}(K_i) \; \forall \, i \neq x \qquad \text{(v)}$$

Substituting (v) into (iv) gives:

$$\Delta s_{\text{in}}(G) = \sum_{i=1 \neq x}^{r} s_{\text{in}}(K_i) + s_{\text{in}}(K_x^*) - \sum_{i=1 \neq x}^{r} s_{\text{in}}(K_i) - s_{\text{in}}(K_x)$$

$$= \; s_{\text{in}}(K_x^*) - s_{\text{in}}(K_x) \qquad \text{(vi)}$$

By definition [D3.1], the change of internal maximum potential number of edges effected by applying $T_z$ to $K_x$ is then:

$$\Delta s_{\text{in}}(K_x) = s_{\text{in}}(K_x^*) - s_{\text{in}}(K_x) \qquad \text{(vii)}$$

Substituting (vii) into (vi) gives:

$$\Delta s_{\text{in}}(G) = \Delta s_{\text{in}}(K_x)$$

<div align="right"><em>QED</em></div>

## Theorem 3.14.

Given the encapsulated graph $G$ with an $i^{th}$ encapsulated region $K_i$ of $\left| K_i \right|$ nodes and external maximum potential number of edges $s_{ex}(K_i)$, the change of $s_{ex}(K_i)$ when the number of information hidden nodes in $K_i$ changes by $m$ where $m \geq -\left| K_i \right|$ is given by:

$$\Delta s_{\text{ex}}(K_i) = m \left| h(G) \right| - m \left| h(K_i) \right|$$

*Proof:*

Let $T_z$ be the hidden transformation defined in [D3.5]; let $K_i^* = T_z(K_i, m)$ and let $G^* = T_z(G, m)$.

By definition [D3.2], the change of external maximum potential number of edges effected by applying $T_z$ to $K_i$ is then:

$$\Delta s_{\text{ex}}(K_i) = s_{\text{ex}}(K_i^*) - s_{\text{ex}}(K_i) \qquad \text{(i)}$$

By theorem 1.4:

$$s_{\text{ex}}(K_i^*) = \left| K_i^* \right| \left( \left| h(G^*) \right| - \left| h(K_i^*) \right| \right) \qquad \text{(ii)}$$

By theorem 3.5:

$$\left| h(G^*) \right| = \left| h(G) \right| \qquad \text{(iii)}$$

Substituting (iii) into (ii) gives:

$$s_{\text{ex}}(K_i^*) = \left| K_i^* \right| \left( \left| h(G) \right| - \left| h(K_i^*) \right| \right) \qquad \text{(iv)}$$

By definintion of $T_z$:

$$\left| h(K_i^*) \right| = \left| h(K_i) \right| \qquad \text{(v)}$$

Substituting (v) into (iv) gives:

$$s_{\text{ex}}(K_i^*) = \left| K_i^* \right| \left( \left| h(G) \right| - \left| h(K_i) \right| \right) \qquad \text{(vi)}$$

By theorem 3.4:

$$\left| K_i^* \right| = \left| K_i \right| + m \qquad \text{(vii)}$$

Substituting (vii) into (vi) gives:

$$s_{\mathrm{ex}}(K_i^*) = \left|K_i + m\right|\left(\left|h(G)\right| - \left|h(K_i)\right|\right)$$

$$= \left|K_i\right|\left|h(G)\right| - \left|K_i\right|\left|h(K_i)\right| + m\left|h(G)\right| - m\left|h(K_i)\right|$$

$$= \left|K_i\right|\left(\left|h(G)\right| - \left|h(K_i)\right|\right) + m\left|h(G)\right| - m\left|h(K_i)\right| \quad \text{(viii)}$$

But:

$$s_{\mathrm{ex}}(K_i) = \left|K_i\right|\left(\left|h(G)\right| - \left|h(K_i)\right|\right) \quad \text{(ix)}$$

Substituting (ix) into (viii) gives:

$$s_{\mathrm{ex}}(K_i^*) = s_{\mathrm{ex}}(K_i) + m\left|h(G)\right| - m\left|h(K_i)\right|$$

$$s_{\mathrm{ex}}(K_i^*) - s_{\mathrm{ex}}(K_i) = m\left|h(G)\right| - m\left|h(K_i)\right|$$

And therefore by (i):

$$\Delta s_{\mathrm{ex}}(K_i) = m\left|h(G)\right| - m\left|h(K_i)\right|$$

*QED*

## Theorem 3.15.

Given the encapsulated graph $G$ of $n$ nodes with an $i^{th}$ encapsulated region $K_i$ of $\left|K_i\right|$ nodes and external maximum potential number of edges $s_{ex}(K_i)$, the change of external maximum potential number of edges of the entire graph when the number of information hidden nodes in a particular $x^{th}$ encapsulated region $K_x$ changes by $m$ where $m \geq -\left|K_x\right|$ is given by:

$$\Delta s_{\mathrm{ex}}(G) = \Delta s_{\mathrm{ex}}(K_x)$$

*Proof:*

By definintion, the external maximum potential number of edges of $G$ is the sum of the external maximum potential number of edges of all its encapsulated regions:

$$s_{ex}(G) = \sum_{i=1}^{r} s_{\mathrm{ex}}(K_i)$$

$$= \sum_{i=1 \neq x}^{r} s_{\mathrm{ex}}(K_i) + s_{\mathrm{ex}}(K_x) \quad \text{(i)}$$

Let $T_z$ be the hidden transformation defined in [D3.5] and let $K_i^* = T_z(K_i, m)$. Futhermore, let $T_z$ apply to the $x^{th}$ encapsulated region only such that $K_x^* = T_z(K_x, m)$ and $G^* = T_z(x, G, m)$. By definintion:

$$s_{ex}(G^*) = \sum_{i=1}^{r} s_{\mathrm{ex}}(K_i^*)$$

$$= \sum_{i=1 \neq x}^{r} s_{\mathrm{ex}}(K_i^*) + s_{\mathrm{ex}}(K_x^*) \quad \text{(ii)}$$

By definintion [D3.2]:

$$\Delta s_{\mathrm{ex}}(G) = s_{\mathrm{ex}}(G^*) - s_{\mathrm{ex}}(G) \quad \text{(iii)}$$

Substituting (i) and (ii) into (iii) gives:

$$\Delta s_{\mathrm{ex}}(G) = \sum_{i=1 \neq x}^{r} s_{\mathrm{ex}}(K_i^*) + s_{\mathrm{ex}}(K_x^*) - \sum_{i=1 \neq x}^{r} s_{\mathrm{ex}}(K_i) - s_{\mathrm{ex}}(K_x) \quad \text{(iv)}$$

By definintion [D3.2]:

$$\Delta s_{ex}(K_x) = s_{ex}(K_x^*) - s_{ex}(K_x) \qquad (v)$$

Substituting (v) into (iv) gives:

$$s_{ex}(G) = \sum_{i=1 \neq x}^{r} s_{ex}(K_i^*) - \sum_{i=1 \neq x}^{r} s_{ex}(K_i) + \Delta s_{ex}(K_x) \qquad (vi)$$

By theorem 1.4:

$$s_{ex}(K_i^*) = \left| K_i^* \right| \left( \left| h(G^*) \right| - \left| h(K_i^*) \right| \right) \qquad (vii)$$

If we consider $K_i$ where $i \neq x$ then as $T_z$ is applied only to $K_x$ then:

$$\left| K_i^* \right| = \left| K_i \right| \; \forall \, i \neq x \qquad (viii)$$

Substituting (viii) into (vii) gives:

$$s_{ex}(K_i^*) = \left| K_i \right| \left( \left| h(G^*) \right| - \left| h(K_i) \right| \right) \qquad (ix)$$

By theorem 3.5:

$$\left| h(G^*) \right| = \left| h(G) \right| \qquad (x)$$

Substituting (x) into (ix) gives:

$$s_{ex}(K_i^*) = \left| K_i \right| \left( \left| h(G) \right| - \left| h(K_i) \right| \right) \qquad (xi)$$

But by theorem 1.4:

$$s_{ex}(K_i) = \left| K_i \right| \left( \left| h(G) \right| - \left| h(K_i) \right| \right) \qquad (xii)$$

Substituting (xii) into (xi) gives:

$$s_{ex}(K_i^*) = s_{ex}(K_i)$$

As this holds $\forall \, i \neq x$ we can take the sum over all encapsulated regions except $x$:

$$\sum_{i=1 \neq x}^{r} s_{ex}(K_i^*) = \sum_{i=1 \neq x}^{r} s_{ex}(K_i) \qquad (xiii)$$

Substituting (xiii) into (vi) gives:

$$s_{ex}(G) = \Delta s_{ex}(K_x)$$

*QED*

## Theorem 3.16.

Given the encapsulated graph $G$ of $n$ nodes with an $i^{th}$ encapsulated region $K_i$ of $\left| K_i \right|$ nodes and external maximum potential number of edges $s_{ex}(K_i)$, the change of external maximum potential number of edges of the entire graph $s_{ex}(G)$ when the number of information hidden nodes in a particular $x^{th}$ encapsulated region $K_x$ changes by $m$ where $m \geq - \left| K_x \right|$ is given by:

$$\Delta s_{ex}(G) = m \left| h(G) \right| - m \left| h(K_x) \right|$$

*Proof:*

Let $T_z$ be the hidden transformation defined in [D3.5] and let $K_i^* = T_z(K_i, m)$. Futhermore, let $T_z$ apply to the $x^{th}$ encapsulated region only such that $K_x^* = T_z(K_x, m)$ and $G^* = T_z(x, G, m)$. By theorem 3.14, the change of external maximum potential number of edges of $K_x$ by the application of $T_z$ to $K_x$ is given by:

$$\Delta s_{ex}(K_x) = m|h(G)| - m|h(K_x)| \quad \text{(i)}$$

By theorem 3.15, the change of external maximum potential number of edges of the entire graph $G$ by the application of $T_z$ to $K_x$ is the same as the change of external maximum potential number of edges of $K_x$ or:

$$s_{ex}(G) = \Delta s_{ex}(K_x) \quad \text{(ii)}$$

Substituting (i) into (ii) gives:

$$\Delta s_{ex}(G) = m|h(G)| - m|h(K_x)|$$

*QED*

## Theorem 3.17.

Given the encapsulated graph $G$ of $n$ nodes with an $i^{th}$ encapsulated region $K_i$ of $|K_i|$ nodes, the change of maximum potential number of edges of the entire graph $s(G)$ when the number of information hidden nodes in a particular $x^{th}$ encapsulated region $K_x$ changes by $m$ where $m \geq -|K_x|$ is given by:

$$\Delta s(G) = m|h(G)| - m|h(K_x)| + 2m|K_x| + m^2 - m$$

*Proof:*

Let $T_z$ be the hidden transformation defined in [D3.5] and let $K_i^* = T_z(K_i, m)$. Furthermore, let $T_z$ apply to the $x^{th}$ encapsulated region only such that $K_x^* = T_z(K_x, m)$ and $G^* = T_z(x, G, m)$. From theorem 3.13, when the number of information hidden nodes in $K_x$ changes by $m$, the change of internal maximum potential number of edges of the entire graph is given by:

$$\Delta s_{in}(G) = \Delta s_{in}(K_x) \quad \text{(i)}$$

By theorem 3.12, when the number of information hidden nodes in $K_x$ changes by $m$, the change of internal maximum potential number of edges of $K_x$ is given by

$$\Delta s_{in}(K_x) = 2m|K_x| + m^2 - m \quad \text{(ii)}$$

Substituting (ii) into (i) gives:

$$\Delta s_{in}(G) = 2m|K_x| + m^2 - m \quad \text{(iii)}$$

From theorem 3.16, when the number of information hidden nodes in $K_x$ changes by $m$, the change of exteral maximum potential number of edges is given by:

$$\Delta s_{ex}(G) = m|h(G)| - m|h(K_x)| \quad \text{(iv)}$$

By theorem 3.1, the change of maximum potential number of edges of $G$ is given by:

$$\Delta s(G) = \Delta s_{in}(G) + \Delta s_{ex}(G) \quad \text{(v)}$$

Substituting (iii) and (iv) into (v) gives:

$$\Delta s(G) = m|h(G)| - m|h(K_x)| + 2m|K_x| + m^2 - m$$

*QED*

## Theorem 3.18.

Given the encapsulated graph $G$ of $n$ nodes, the cumulative change of maximum potential number of edges of the entire graph $s(G)$ when $m$ information hiding violational nodes are moved from a particular source encapsulated region $K_s$ to a particular target encapsulated region $K_t$ is given by:

$$\Delta s_{cumulative}(G) = m(|K_t| - |h(K_t)| - (|K_s| - |h(K_s)|))$$

*Proof:*

Let $T_{h1}$ be the violational transformation defined in [D3.4] and let $T_{h1}$ apply to the $s^{th}$ encapsulated region $K_s$ only such that $K_s^* = T_{h1}(K_s, m)$ and $G^* = T_{h1}(s, G, m)$. $T_{h1}$ will remove $m$ information hiding violational nodes from $K_s$.

Let $T_{h2}$ be the violational transformation defined in [D3.4] and let $T_{h2}$ apply to the $t^{th}$ encapsulated region $K_t$ only such that $K_t^* = T_{h2}(K_t, m)$ and $G^{**} = T_{h2}(t, G^*, m)$. $T_{h2}$ will add $m$ information hiding violational nodes from $K_t$.

Let us define the violational translation transformation $T_{Th}$ as the combination of the two translations $T_{h1}$ and $T_{h2}$ such that:

$$T_{Th}(G) = T_{h2}(T_{h1}(G))$$

As these transformations are linear, the change of the maximum potential number of edges of $T_{Th}$ is equal to the sum of the changes of the maximum potential number of edges of the transformations $T_{h2}$ and $T_{h1}$, or:

$$\Delta s_{cumulative}(G) = \Delta s(G) + \Delta s(G^*) \quad \text{(i)}$$

Considering nodes removed from an encapsulated region as negative, then $T_{h1}$ will add $-m$ nodes to $G$. By theorem 3.11, the change of maximum potential number of edges in $G$ caused by the removal of these $m$ information hiding violational nodes from $K_s$ is given by:

$$\Delta s(G) = mn + m|K_s| + m|h(G)| - m|h(K_s)| + m^2 - m \quad \text{(ii)}$$

Substituting $-m$ for $m$ in (ii) gives:

$$\Delta s(G) = -mn - m|K_s| - m|h(G)| + m|h(K_s)| + m^2 + m \quad \text{(iii)}$$

By theorem 3.11 again, the change of maximum potential number of edges in $G^*$ caused by the addition of $m$ information hiding violational nodes to $K_t$ is given by:

$$\Delta s(G^*) = mn^* + m|K_t| + m|h(G^*)| - m|h(K_t)| + m^2 - m \quad \text{(iv)}$$

$G^*$ has $m$ fewer nodes than $G$ and they are all information hiding violational, and thus both $h(G^*)$ and $n^*$ are changed in comparison with $G$, such that:

$$|h(G^*)| = |h(G)| - m \quad \text{(v)}$$

$$n^* = n - m \quad \text{(vi)}$$

Substituting (v) and (vi) into (iv) gives:

$$\Delta s(G^*) = m(n - m) + m|K_t| + m(|h(G)| - m) - m|h(K_t)| + m^2 - m$$

$$= mn - m^2 + m|K_t| + m|h(G)| - m^2 - m|h(K_t)| + m^2 - m$$

$$= mn + m|K_t| + m|h(G)| - m|h(K_t)| - m^2 - m \quad \text{(vii)}$$

Substituting (iii) and (vii) into (i) gives:

$$\Delta s_{cumulative}(G) = -mn - m|K_s| - m|h(G)| + m|h(K_s)| + m^2 + m$$
$$+ mn + m|K_t| + m|h(G)| - m|h(K_t)| - m^2 - m$$

$$= m\left(\left|K_t\right|-\left|h\left(K_t\right)\right|-\left(\left|K_s\right|-\left|h\left(K_s\right)\right|\right)\right)$$

<div align="right"><em>QED</em></div>

## Theorem 3.19.

Given the encapsulated graph $G$ of $n$ nodes, the cumulative change of maximum potential number of edges of the entire graph $s(G)$ when $m$ information hidden nodes are moved from a particular source encapsulated region $K_s$ to a particular target encapsulated region $K_t$ is given by:

$$\Delta s_{cumulative}(G)=m\left(2\left|K_t\right|-2\left|K_s\right|+\left|h\left(K_s\right)\right|-\left|h\left(K_t\right)\right|+2m\right)$$

*Proof:*

Let $T_{z1}$ be the hidden transformation defined in [D3.5] and let $T_{z1}$ apply to the $s^{th}$ encapsulated region $K_s$ only such that $K_s{}^*=T_{z1}(K_s,m)$ and $G^*=T_{z1}(s,G,m)$. $T_{z1}$ will remove $m$ information hidden nodes from $K_s$.

Let $T_{z2}$ be the hidden transformation defined in [D3.5] and let $T_{z2}$ apply to the $t^{th}$ encapsulated region $K_t$ only such that $K_t{}^*=T_{z2}(K_t,m)$ and $G^{**}=T_{z2}(t,G^*,m)$. $T_{z2}$ will add $m$ information hidden nodes from $K_t$.

Let us define the hidden translation transformation $T_T$ as the combination of the two translations $T_{z1}$ and $T_{z2}$ such that:

$$T_T(G)=T_{z2}(T_{z1}(G))$$

As these transformations are linear, the change of the maximum potential number of edges of $T_T$ is equal to the sum of the changes of the maximum potential number of edges of the transformations $T_{z2}$ and $T_{z1}$, or:

$$\Delta s_{cumulative}(G)=\Delta s(G)+\Delta s(G^*) \quad \text{(i)}$$

Considering nodes removed from an encapsulated region as negative, then $T_{z1}$ will add $-m$ nodes to $G$. By theorem 3.17, the change of maximum potential number of edges in $G$ caused by the removal of these $m$ information hidden nodes from $K_s$ is given by:

$$\Delta s(G)=m\left|h(G)\right|-m\left|h(K_s)\right|+2m\left|K_s\right|+m^2-m \quad \text{(ii)}$$

Substituting $-m$ for $m$ in (ii) gives:

$$\Delta s(G)=-m\left|h(G)\right|+m\left|h(K_s)\right|-2m\left|K_s\right|+m^2+m \quad \text{(iii)}$$

By theorem 3.17 again, the change of maximum potential number of edges in $G^*$ caused by the addition of $m$ information hidden nodes to $K_t$ is given by:

$$\Delta s(G^*)=m\left|h(G^*)\right|-m\left|h(K_t)\right|+2m\left|K_t\right|+m^2-m \quad \text{(iv)}$$

$G^*$ has m fewer nodes than $G$ but they are all information hidden, and thus $h(G^*)$ is unchanged, or:

$$\left|h(G^*)\right|=\left|h(G)\right| \quad \text{(v)}$$

Substituting (v) into (iv) gives:

$$\Delta s(G^*)=m\left|h(G)\right|-m\left|h(K_t)\right|+2m\left|K_t\right|+m^2-m \quad \text{(vi)}$$

Substituting (iii) and (vi) into (i) gives:

$$\Delta s_{cumulative}(G)=-m\left|h(G)\right|+m\left|h(K_s)\right|-2m\left|K_s\right|+m^2+m$$
$$+m\left|h(G)\right|-m\left|h(K_t)\right|+2m\left|K_t\right|+m^2-m$$
$$= m\left(2\left|K_t\right|-2\left|K_s\right|+\left|h(K_s)\right|-\left|h(K_t)\right|+2m\right)$$

## Theorem 3.20.

Given the encapsulated graph $G$ of $n$ nodes, the cumulative change of maximum potential number of edges of the entire graph $s(G)$ when $m$ information hidden nodes in a particular encapsulated region $K_x$ are converted to information hiding violational nodes within the same encapsulated region is given by:

$$\Delta s_{cumulative}(G) = m(n - |K_x|)$$

*Proof:*

Let $T_z$ be the hidden transformation defined in [D3.5] and let $T_z$ apply to the $x^{th}$ encapsulated region $K_x$ only such that $K_x^* = T_z(K_x, m)$ and $G^* = T_z(x, G, m)$. $T_z$ will remove $m$ information hidden nodes from $K_x$.

Let $T_h$ be the violational transformation defined in [D3.4] and let $T_h$ apply to the $x^{th}$ encapsulated region $K_x$ only such that $K_x^* = T_h(K_x, m)$ and $G^{**} = T_h(x, G^*, m)$. $T_h$ will add $m$ information hiding violational nodes from $K_x$.

Let us define the conversion transformation $T_C$ as the combination of the two translations $T_z$ and $T_h$ such that:

$$T_C(G) = T_h(T_z(G))$$

As these transformations are linear, the change of the maximum potential number of edges of $T_C$ is equal to the sum of the changes of the maximum potential number of edges of the transformations $T_h$ and $T_z$, or:

$$\Delta s_{cumulative}(G) = \Delta s(G) + \Delta s(G^*) \quad \text{(i)}$$

Considering nodes removed from an encapsulated region as negative, then $T_z$ will add $-m$ nodes to $G$. By theorem 3.17, the change of maximum potential number of edges in $G$ caused by the removal of these $m$ information hidden nodes from $K_x$ is given by:

$$\Delta s(G) = m|h(G)| - m|h(K_x)| + 2m|K_x| + m^2 - m \quad \text{(ii)}$$

Substituting $-m$ for $m$ in (ii) gives:

$$\Delta s(G) = -m|h(G)| + m|h(K_x)| - 2m|K_x| + m^2 + m \quad \text{(iii)}$$

By theorem 3.11, the change of maximum potential number of edges in $G^*$ caused by the addition of $m$ information hiding violational nodes to $K_x$ is given by:

$$\Delta s(G^*) = mn^* + m|K_x^*| + m|h(G^*)| - m|h(K_x^*)| + m^2 - m \quad \text{(iv)}$$

$K_x^*$ has $m$ fewer nodes than $K$ and they are all information hidden, and thus both $K_x^*$ and $n^*$ are changed in comparison with $G$, but $h(K_x^*)$ and $h(G^*)$ are unchanged such that:

$$|h(G^*)| = |h(G)| \quad \text{(v)}$$

$$|h(K_x^*)| = |h(K_x)| \quad \text{(vi)}$$

$$n^* = n - m \quad \text{(vii)}$$

$$|K_x^*| = |K_x| - m \quad \text{(viii)}$$

Substituting (v), (vi), (vii) and (viii) into (iv) gives:

$$\Delta s(G^*) = m(n - m) + m(|K_x| - m) + m|h(G)| - m|h(K_x)| + m^2 - m$$

$$= \quad mn - m^2 + m|K_x| - m^2 + m|h(G)| - m|h(K_x)| + m^2 - m$$

$$= \quad mn + m|K_x| + m|h(G)| - m|h(K_x)| - m^2 - m \quad \text{(ix)}$$

Substituting (iii) and (ix) into (i) gives:

$$\Delta s_{cumulative}(G) = -m|h(G)| + m|h(K_x)| - 2m|K_x| + m^2 + m$$
$$+ mn + m|K_x| + m|h(G)| - m|h(K_x)| - m^2 - m$$

$$= \quad m(n - |K_x|)$$

*QED*